\documentclass[runningheads]{svmult}

\newcommand{\lsim}{\, \lower2truept\hbox{${<
\atop\hbox{\raise4truept\hbox{$\sim$}}}$}\,}
\newcommand{\gsim}{\, \lower2truept\hbox{${>
\atop\hbox{\raise4truept\hbox{$\sim$}}}$}\,}

\usepackage{makeidx}   
\usepackage{graphicx}  

\usepackage{multicol}  
\usepackage{physprbb}  
\makeindex             

\begin{document}
\title*{Cold Dark Matter Halos Must Burn }
%
%

%
%
\titlerunning{Dark Halos and CDM}
%

\author{Paolo Salucci \inst{1}
\and Annamaria Borriello \inst{1} }
\authorrunning{Salucci \& Borriello}
\institute{(1) International School for Advanced Studies  SISSA Trieste,  Italy}

\maketitle              

\begin{abstract}
High--quality  optical rotation curves  for  a sample of low--luminosi\-ty 
spirals   evidence  that the dark halos around galaxies are inconsistent with
the  output of  proper CDM  simulations. In fact, dark   halos enveloping 
stellar disks are structures with approximately a constant density out to the
optical edges. This is in strong disagreement  with  the characteristic
$\rho(r) \propto r^{-1.5}$ CDM regime and  severely  challenges  the
``standard"  CDM  theory also because the  halo density appears to be  heated
up, at gross variance with the hierarchical evolution of collision--free
particles.   \end{abstract}

\section{CDM and Galaxy Halos}

Dark matter (DM) halos  in  the  Cold Dark Matter  scenario  are  formed  via
dissipationless  hierarchical merging  and harbor the infall/cooling of
the   primordial gas that leads  to the formation of  the
present--day galaxies. Several studies have investigated the
detailed structure of these halos  by means of   $N$-body simulations at a
progressively higher  spatial/mass  resolution (e.g.  \cite{NFW96}, \cite{Ma99}).
The outcome is well known:    CDM halos have an ``universal"  profile    in    their
density which includes  a steep central cusp.    In detail,    on galactic scales
$\simeq 1-10 $  kpc:  $\rho(r)  \propto   r^{-\gamma}$. Simulations
at the highest resolution  indicate   $\gamma = -1.5$
\cite{Ga00}.
 \begin{table}
\begin{center}
\begin{tabular}{lccccl}
\hline
\hline
\noalign{\smallskip}
Name  &    $M_I$    & $R_{\rm opt}$  &   $\beta$  &  $\rho_s $  &   $r_s  $   \\
\hline
\noalign{\smallskip}
 116-G12         &  $-20.0$  &    5.4	& $0.29^{+0.03}_{-0.1}$  &
$2.7^{+3}_{-0.7}$  &  $10^{+10}_{-5}$	\\
 531-G22         &  $-21.4$  &     10.5	&  $0.11^{+0.07}_{-0.06}$  &
$2.1^{+0.5}_{-0.4}$  &  $12^{+4}_{-2}$	\\
533-G4           &  $-20.7$   &    8.4	& $0.07^{+0.05}_{-0.02}$  &
$4.3^{+0.5}_{-0.9}$  &  $6^{+2}_{-1}$	\\
545-G5           &  $-20.4$   &    7.5	& $0.21^{+0.03}_{-0.03}$  &
$1.0^{+0.3}_{-0.3}$  &  $22^{+300}_{-4}$ 	\\
563-G14         &  $-20.5$   &    6.3	& $0.25^{+0.02}_{-0.07}$  &
$4.3^{+1}_{-0.9}$  &  $6^{+2}_{-2}$ 	\\
79-G14            &  $-21.4$  &    12.3 	&$0.19^{+0.03}_{-0.07}$  &
$1.3^{+0.3}_{-0.3}$  &  $14^{+4}_{-2}$	\\
M-3-1042       &  $-20.1$  &    4.7 	&$0.45^{+0.06}_{-0.04}$  &
$2.2^{+3}_{-0.8}$  &  $15^{+300}_{-10}$	\\
N7339              &  $-20.6$  &    4.9 	& $0.53^{+0.05}_{-0.03}$  &
$2.1^{+0.3}_{-0.5}$  &  $22^{+300}_{-7}$	\\
N755                &  $-20.1$   &    4.8	& $0.05^{+0.02}_{-0.04}$  &
$4.1^{+1}_{-0.7}$  &  $5^{+3}_{-1}$  	\\
\hline
\end{tabular}
\end{center}
\end{table}

\begin{figure}[h]
\vspace{-1.3truecm} \hspace{-2.5truecm}
\includegraphics[width=165mm,height=160mm]{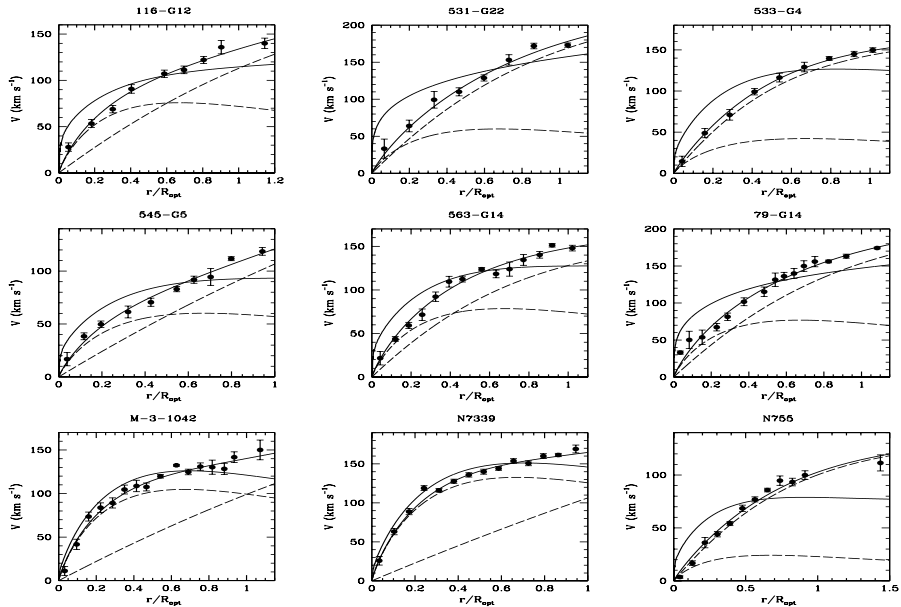}
\vspace{-6.2truecm}
\caption[]{  RC's  of the sample ({\it points})
{\it vs}     CDM  model ({\it  discrepant
continuous line}) and {\it vs}  BS  model,  of which    the disk  and  halo
contributions are shown  as  ({\it dotted lines}).}
\label{eps1}
\end{figure}

Let us  represent   generic  halos,   including the  CDM  ones,  by means of:
\begin{equation}
\rho(r) = {\rho_s  \over [c+(r/r_s)^{\gamma}] \
 [1+(r/r_s)^{\alpha}]
^{(\delta-\gamma)/\alpha}}
\end{equation}
where  $\gamma$ is the density inner slope,  $\delta$ is  the outer    slope and
$\alpha$ settles  the turnover   point between the inner and outer regimes. The
parameter  $c$ indicates the presence of  a   ``constant density"  (inner) region
(CDR).  If $c=0$, as   in collisionless  CDM, the  density diverges for $r
\rightarrow 0$.  In the case  of $c=1$,   $r_s$ is the
size of  the  CDR,  whose  value is  $\rho_s$.  According to the above  notation,
the Moore density profile  corresponds  to ($c$,  $\alpha$, $\delta$,
$\gamma$)=  (0, 1.5, 3, 1.5) and the Burkert--Salucci one  to  (1, 2, 3, 1)  
\cite{B95}.

We test these mass models  by means of  the  dark halos  detected around
galaxies of  a sample of  low luminosity late--type spirals.   In these
objects the   DM distribution can be easily derived from  the available
high-quality {\it optical} rotation curves because   the stellar  and HI  disks
contribute very little  to the gravitational potential so as
the beam smearing effect, plaguing  the HI RC's, is obviously absent.
The sample  includes  9 high--quality   {\it optical} RC's that   are
smooth and symmetric and  extend  at least out  to  the optical
edge $R_{opt}$. The  characteristics of merit are the following: the  spatial
resolution is   better than $1/20\ R_{\rm opt}$ and  the  velocity  {\it  rms} is
$<3\%$. The  mass modeling is furthermore simplified in that  in these spirals  the
gas contribution  to $V(r)$  (for $R<R_{\rm opt}$) is  modest  \cite{BS01} and
the stellar surface brightness  is very well fit by a Freeman disk.
\begin{figure}[h] \vspace{-1.5truecm} \hspace{-1.7truecm} 
\includegraphics[width=150mm,height=205mm]{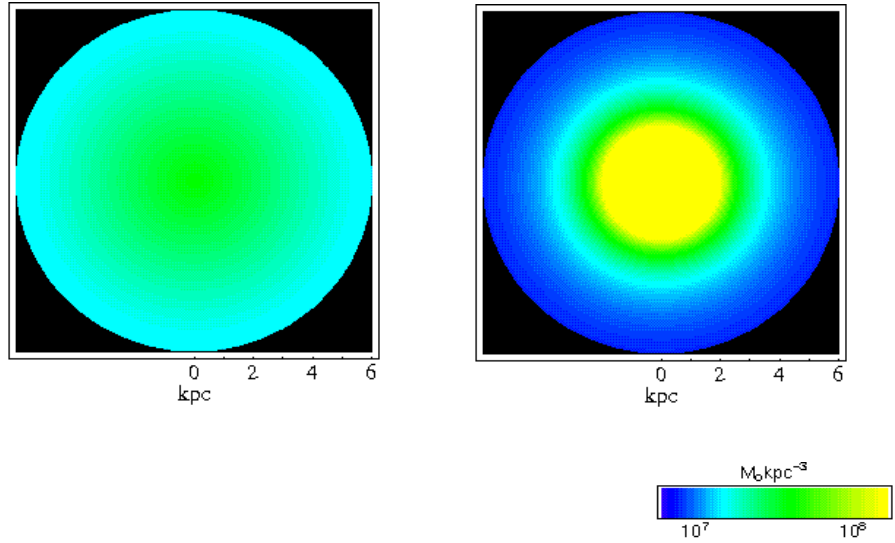}
\vspace{-11.3truecm}
\caption[]{{\it left}) Dark halo around  116--G12, {\it right}) CDM
prediction for a halo of the same mass (inside $R_{\rm opt}) $.}
\label{eps2}
\end{figure}

\section{The Structure of Dark  Halos }

The  mass model includes {\it i)} a stellar exponential thin disk  with free
parameter  $\beta \equiv  (V_D  /   V)^2_{R_{\rm opt}}$ i.e.     the fractional
disk contribution  to the total circular velocity at $R_{\rm opt}$,  and {\it ii)} a
dark spherical halo, whose  contribution to $V(r)$  is  obtained from (1) in that:
$V^2_h(r) \equiv   \frac{G}{r} \int_0^r 4 \pi r^2 \rho(r)  dr$. Once we choose the
scenario (i.e. $c=1$ or   $ c=0$),  we are left with  two free parameters,
$\rho_s$ and $r_s$, the characteristic  density and  radius. Both   get
determined by adjusting the model to data as in   \cite{BS01}.

The CDM halo  profiles  fail:   in no case they  can reproduce   (with or without an
exponential disk) the  observed RC's.  This is  shown in Fig. 1: the discrepancy
with data is very high at any radius. The  CDM halos, even allowed  to take any
value for $r_s$ (or for the concentration parameter), have   definitely a  too steep
density profile  in the   innermost region and show a  too flat profile  in the
outer regions.

Next, we  fit  the BS   profile  to the data: the results   are  shown in Fig. 1
and Table 1.  This model fits  perfectly, indicating the values of  parameters
which are  reported in Table 1. Each of 9 halos has a  central density $\rho_s$ of
about $1-4 \times   10^{-24}$ g/cm$^3$ that keeps constant  out the edge of the
stellar distribution.

\section{You  Don't Need More Evidence}

We   definitely  conclude, as in \cite {BS01},  that
the dark halos embedding the stellar disks  show a density distribution inconsistent
with that predicted by CDM (see  Fig.2).  Real halos in the Universe  rather
resemble, in the regions where  the stars reside,  some homogeneous spheres.
Crucially,  the sample and the method employed here level off  to zero the
criticisms raised to previous claims for DM  core  radii in galaxies.
Furthermore,   since the  DM  radial  distribution out to $R_{opt}$
is featureless,  we cannot  link together the local and the   global properties of
the dark halos; such a  connection, which is a main consequence  of the bottom--up
merging  scenario appears to be just  missing  in Nature. The CDM  scenario, then,
must find a way to cut off any  signature of the  gravitational
collisionless collapse. For instance,  the ``temperature" of the CDM particles, which
initially $\rightarrow 0$  for  $r \rightarrow 0$, must be largely  heated up and kept  
constant  out to the disk edge  (see \cite{bur2}).
Solving  this deep mystery will be the guideline of our (and others) future research.

{}

\end{document}